\documentclass{sig-alternate}
\usepackage{mathptmx} 

\usepackage{fancyhdr}
\usepackage[normalem]{ulem}
\usepackage[hyphens]{url}
\usepackage[sort,nocompress]{cite}
\usepackage[final]{microtype}
\usepackage[keeplastbox]{flushend}
\usepackage[bookmarks=true,breaklinks=true,letterpaper=true,colorlinks,citecolor=blue,linkcolor=blue,urlcolor=blue]{hyperref}

\usepackage{amsmath,amsfonts}
\usepackage{array}

\usepackage[caption=false,font=normalsize,labelfont=sf,textfont=sf]{subfig}
\usepackage{stfloats}
\usepackage{verbatim}
\usepackage{graphicx}
\usepackage{cite}
\usepackage{listings}
\usepackage[table]{xcolor}
\usepackage{tcolorbox}
\usepackage{blindtext}
\usepackage{tikz}
\usepackage{multicol}
\usepackage{wrapfig}
\usepackage{kantlipsum}
\usepackage{boldline}
\usepackage{float}
\usepackage{xspace}
\usepackage{booktabs, makecell, multirow, tabularx}
\usepackage{todonotes} 

\newcommand{\ifmap}{\textit{ifmap}\xspace}
\newcommand{\ofmap}{\textit{ofmap}\xspace}
\newcommand{\ifmaps}{\textit{ifmaps}\xspace}
\newcommand{\ofmaps}{\textit{ofmaps}\xspace}
\newcommand{\fmap}{\textit{fmap}\xspace}

\newcommand{\T}{\triangleright}

\definecolor{dkgreen}{rgb}{0,0.6,0}
\definecolor{gray}{rgb}{0.5,0.5,0.5}
\definecolor{mauve}{rgb}{0.58,0,0.82}

\newcommand*\circled[1]{\tikz[baseline=(char.base)]{
            \node[shape=circle,fill,inner sep=0.3pt] (char) {\textcolor{white}{#1}};}}

\newcommand*\circlenew[1]{\tikz[baseline=(char.base)]{
            \node[shape=circle,draw,inner sep=0.3pt] (char) {#1};}}      
\newcommand{\fname}{{\em Seculator}\xspace}

\title{Seculator: A Fast and Secure Neural Processing Unit} 
\numberofauthors{2} 
\author{ \alignauthor Nivedita Shrivastava \\ \affaddr {Electrical Engg., IIT Delhi} \\ \affaddr{New Delhi, India} \\
\email{nivedita.shrivastava@ee.iitd.ac.in}  \alignauthor Smruti R. Sarangi \\ \affaddr{Electrical Engg., IIT Delhi} \\
\affaddr{New Delhi, India} \\ \email{srsarangi@cse.iitd.ac.in} }

\begin{document}
\maketitle
\pagestyle{plain}

\begin{abstract}
Securing deep neural networks (DNNs) is a problem of significant interest since an ML model incorporates high-quality intellectual
property, features of data sets painstakingly collated by mechanical turks, and novel methods of training on large
cluster computers. Sadly, attacks to extract model parameters are on the rise, and thus designers are being forced to
create architectures for securing such models. State-of-the-art proposals in this field take the deterministic memory access
patterns of such networks into cognizance (albeit partially), group a set of memory blocks into a 
{\em tile}, and maintain state at the level of tiles (to reduce storage space). 
For providing integrity guarantees (tamper avoidance), they don't propose
any significant optimizations, and still  maintain block-level state.    

We observe that it is possible to exploit the deterministic memory access patterns of DNNs even further, and 
maintain state information for only the current tile and current layer, 
which may comprise a large number of tiles. This reduces
the storage space, reduces the number of memory accesses, increases performance, and simplifies the design without
sacrificing any security guarantees.
The key techniques in our
proposed accelerator architecture, \fname, are to
encode memory access patterns to create a small HW-based tile version number generator for a given layer, 
and to store layer-level MACs. 
We completely eliminate the need for having a MAC cache and a tile version number store (as used in related work). 
We show that using intelligently-designed mathematical operations, these structures are not required. By reducing such
overheads, we show a speedup of
16\% over the closest competing work. 
\end{abstract}

\section{Introduction}

The AI hardware (Neural Processing Unit (NPU)) market was valued at 8 billion USD in 2020 and is expected to grow to 
84 billion USD by 2028~\cite{verifiedmarketresearch} (CAGR of 34.15\%). Hardware-based
AI chips are expected to play a major role in telecommunications~\cite{facon2019hardware}, mobile
vision~\cite{vision}, edge computing~\cite{edge}, augmented/virtual reality~\cite{AR}, 
health care~\cite{healthcare}, and autonomous driving~\cite{vehicle}. Similar to code for general-purpose
processors, ML models
incorporate a lot of high-value intellectual property (IP) that takes a lot of time and
effort to collate and develop. For example, even for a mid-size AI project just collecting
the raw data using mechanical turks takes upwards of \$100K USD~\cite{MLcost}; the know-how 
for the model and training methodology can be worth millions of dollars, and finally 
it may take many weeks to finally train the model using large cluster computers.
Along with these conventional arguments, many a time we don't realize that even getting access
to the training data can be quite challenging with numerous legal hurdles. 
Therefore, protecting an ML model is of paramount importance, which sadly also makes it 
an attractive target for hackers.

\begin{figure}[!h]
    \centering
    \includegraphics[scale=0.55]{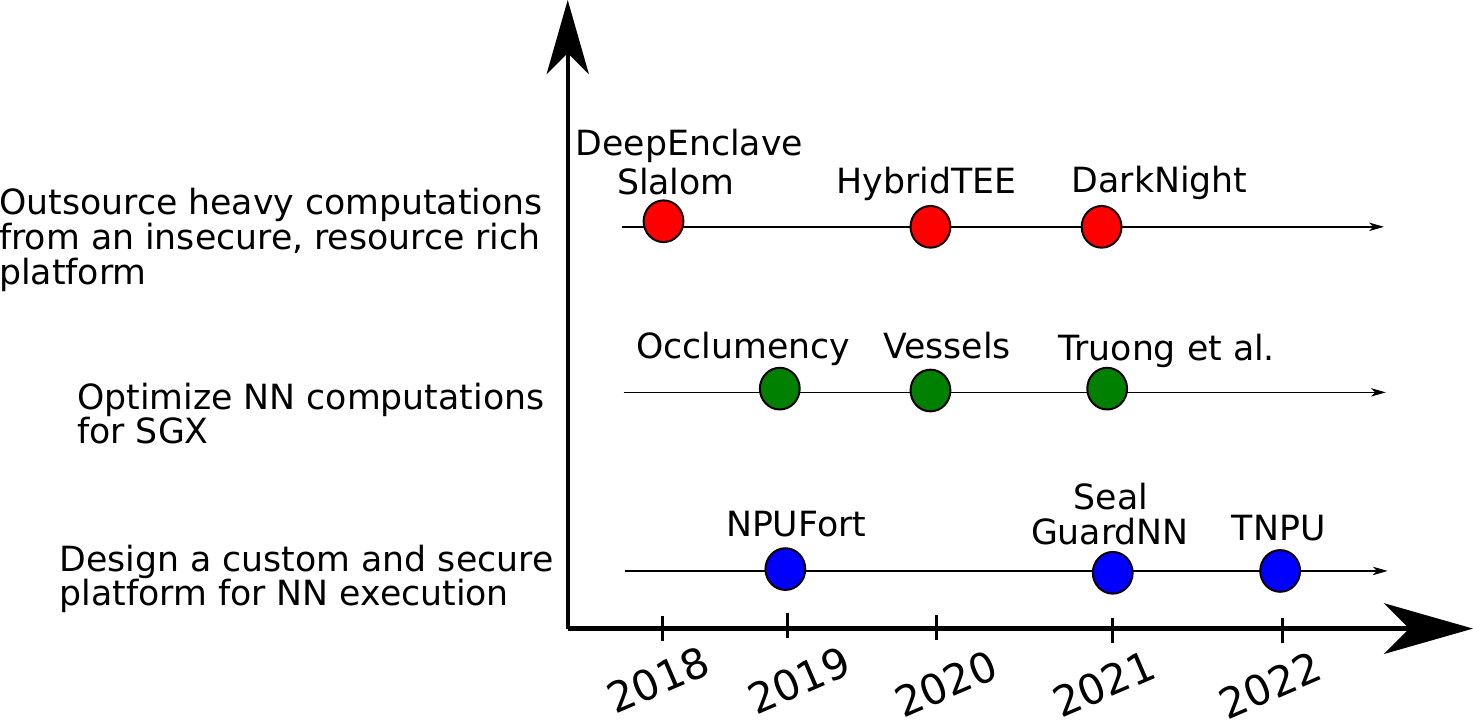}
    \caption{A timeline depicting a pragmatic shift from outsourcing execution from an 
unsecure compute-rich platform to designing an optimized and secure custom TEE for accelerating ML workloads.}
    \label{fig:timeline}
\end{figure}

There are three broad approaches for protecting models as shown in Figure~\ref{fig:timeline}. All of them
rely on secure CPUs as the baseline technology. To secure CPUs, we rely on Trusted Execution 
Environments (TEEs) such as Intel SGX~\cite{intelsgx}, AMD SEV~\cite{amd}, and ARM Trustzone~\cite{trustzone}.
In all these TEEs, the CPU is assumed to be secure; it is within the {\em TCB} (Trusted 
Computing Base). The main contribution of the most elaborate scheme, Intel SGX, 
is in protecting the off-chip memory and providing
confidentiality, integrity, and freshness. The key insight is ensuring that every time 
a block is written to main memory, it is encrypted with a different key. Since, storing
a key for every memory block is too expensive, a more efficient mechanism is used based on
AES counter-mode encryption. Every page is associated with a major counter, and every block
uses a minor counter (a combination of both guides the encryption/decryption). The counters
themselves need to be protected; this is achieved using a Merkle tree, where the root of 
the tree is guaranteed to be in the TCB.  Let us refer to this version of SGX as {\em SGX-Client}.
Because of the Merkle tree and associated overheads of maintaining counters, the maximum 
size of the protected memory region is limited to 128-256 MB (same problem with
ARM TrustZone). Most ML models and datasets
as of today  are much larger. As a result, {\em SGX-Client} is not the best choice for them.

Keeping in mind these issues, Intel recently discontinued support for {\em SGX-Client} in its
$11^{th}$ and $12^{th}$ generation processors. It replaced it with another version of SGX 
(referred to as {\em SGX-Server}) that simply encrypts memory and foregoes the integrity and
freshness guarantees~\cite{xeon} (on the lines of AMD SEV). It can provide up to 512 GB of encrypted
memory. We should bear in mind that {\em SGX-Server} is far weaker than {\em SGX-Client} in
terms of the security guarantees that are provided namely integrity and freshness.

ML architecture designers have traditionally opted to use versions of {\em SGX-Client} to secure
their systems (refer to Figure~\ref{fig:timeline}). Several early approaches either proposed optimizations
to {\em SGX-Client}~\cite{xeon,intelsgx} to reduce its overhead or partition an ML algorithm into a secure portion
and unsecure portion that ran on fast hardware such as GPUs~\cite{slalom}. However, these approaches have
been superseded by a newer family of approaches that leverage the stable data communication
patterns of ML workloads to design bespoke NoCs. Two custom accelerators stand out in this space: 
TNPU~\cite{tnpu} and GuardNN~\cite{guardnn}. Their basic ideas are similar: group a set of contiguous memory
blocks into {\em tiles} and provide freshness guarantees at the level of a tile. Both use a dedicated
software module running on the host CPU
for managing version numbers (VNs); they are used to encrypt data as well as ensure its freshness. 
The major difference is that TNPU stores the VNs in an on-chip cache and
GuardNN relies on a CPU program to generate all the VNs. The program needs to store all the VNs
in use in secure memory, and securely communicate them to the accelerator (a difficult problem
in itself).
Both still perform integrity checking at the 
level of individual memory blocks (typically 64 bytes).

Our scheme, \fname, improves upon these ideas and proposes a natural extension, where we 
perform freshness and integrity checks at
\underline{the layer level}. Given that no data is stored and no checking is done at the block-level 
(like previous schemes), there is an associated performance gain (15.6\%). Additionally, there is
no need to run a VN-generation module on the host CPU using a TEE. 
To realize this, we thoroughly characterize the memory access patterns of different kinds of ML accelerators, efficiently
encode them, and pass the encoding to a small hardware circuit that automatically generates all VNs
at runtime (without external intervention). This eliminates the need for storing and managing VNs in on-chip 
caches or main memory regions. Furthermore, we leverage the insight that the only consumer for the output(s) of 
a layer are a few layers that it is connected to -- random access of output data is not required because these
consumers layers access the data in a structured fashion. Hence, computing and storing
MACs at the layer-level is good enough as long as the next layer accesses all the data (in any order), which
is often the case.

To summarize, {\bf the main contributions in this paper} are as follows: \circlenew{1} Characterization of traffic in 
CNNs and popular data pre/post processing algorithms, \circlenew{2} A method to succinctly encode the traffic pattern,
\circlenew{3} A method to generate VNs on the fly using such patterns, \circlenew{4} A technique to
 perform integrity checking at the level of layers, and \circlenew{5} A detailed experimental analysis of \fname that
shows a 16\% speedup over the nearest competing work. \circlenew{6} Given that the overheads are low, we additionally
assess the benefits of interspersing the execution with the running of a dummy network (for the purpose of
adding noise) or widening each layer. This helps reduce the possibility of model extraction 
attacks by utilizing timing or address
side channels substantially.

The paper is organized as follows.
Section~\ref{sec:background} presents the background of TEEs and basic convolution
operations in neural networks. Section~\ref{sec:ThreatModel} presents the threat model and the 
security guarantees provided by \fname.
We characterize the workloads in Section~\ref{sec:characterization}, 
present an analytical model for automatic version number generation in   Section~\ref{sec:PatternAnalysis}, and
present the architecture of \fname in 
Section \ref{sec:architecture}. Finally, Section \ref{sec:results} reports the
experimental results, Section~\ref{sec:RelatedWork} presents the related work and we conclude in Section
\ref{sec:conclusion}.

\section{Background}\label{sec:background}
 A TEE provides an isolated execution environment for security-sensitive applications, ensuring data confidentiality and
integrity. The security community has relied on them for performing secure data computations even in an insecure
environment.

\subsection{Intel SGX}
SGX is a TEE provided by Intel to enable remote code execution in a \textit{secure enclave} by isolating
security-sensitive code and data from other applications. These hardware-assisted secure enclaves are designed to handle
computations securely, even if the operating system (OS) and the hypervisor are untrusted. 

\subsubsection{SGX-Client ($10^{th}$ Gen Intel CPUs)}
\label{sec:sgxclient}
\textit{SGX-Client} provides a protected memory of limited size, $256$ MB, which is accessible from within the secure
enclave only. If the data size exceeds this limit, there will be a significant overhead due to pages' eviction and
encryption. 

\textit{Confidentiality With Memory Encryption: }
SGX uses \textit{AES-CTR} (counter mode) for performing encryption. A data block $P$ is XORed with a one-time pad, which
is generated by encrypting the address of the data block $PA$ and the counter value $C$ (major+minor counter) 
with a secret key. The secret
key is a concatenation of the enclave ID $E$, PUF $P$, and a random number generated at boot time $R$. The entire
process can be written as: $P \oplus AES_{(E||P||R)}(PA||C)$ , where $||$ represents a concatenation operator and
$\oplus$ represents a XOR operator. Each block in a page is assigned a 6-bit minor counter, which is concatenated with a
64-bit page-level major counter.  The counter value is incremented when a modified cache block is evicted from the
last-level cache (LLC)~\cite{book}. The major counter is incremented once a minor counter overflows.
These steps ensure that the combined counter value $C$ is never reused, thus avoiding a
\textit{replay attack}. 
The counter values are stored inside a secure cache known as a \textit{counter-cache}. If an entry is not present in
the cache, it is fetched from main memory. A Merkle tree guarantees the integrity of the counters stored in DRAM. 

\textit{Integrity and Authenticity Verification using Message Authentication Codes (MACs): } MACs are encrypted hashes,
that are used to ensure that an adversary does not alter the data stored in an untrusted location. A unique MAC is
generated for each data block and the corresponding counter value. When a data block is fetched from main memory, its
MAC is also fetched and verified. As the data encryption takes into account the block address, an attacker cannot swap
the $\langle data, MAC \rangle$ pair with another pair. Sadly, these security guarantees come with significant
overheads, and thus the size of protected memory is limited to 256 MB.

\subsubsection{SGX-Server ($11^{th}$ and $12^{th}$ Gen Intel CPUs)}
Intel recently launched a scalable and efficient TEE~\cite{xeon} for Intel $3^{rd}$ Gen scalable Xeon
servers with an additional feature, \textit{Total Memory
Encryption (TME)}, to overcome the size limitations of SGX-Client. 
This feature provides a secure memory size of 512 GB. 
Sadly, Intel compromised hardware-based integrity and replay protection in order to securely
encrypt the entire memory (as mentioned in their documentation~\cite{xeon,tnpu}). 
SGX-Server uses AES-XTS for performing total memory encryption, which does
not rely on per-block counters. With the elimination of counters and the Merkle tree, the need for data caching and tree
traversal are also eliminated, thus reducing the associated overheads.

\subsection{Convolution}
A convolution operation is the heart of a convolutional neural network (CNN) and many other modern ML algorithms such as LSTMs,
RNNs, and GANs. The single-image version has a basic 6-loop structure. We iterate through the input and output feature
maps (\ifmap and \ofmap, resp.), and compute the convolutions. For the ease of explanation, we assume that
they have the same size (both are referred to as an \fmap) and are 2D matrices (each element is a {\em pixel}).

\begin{lstlisting} [caption={A basic convolution operation},captionpos=b,label=lst:conv]
for(k=0; k<K ; k++){      // K: #output fmaps
 for(c=0; c<C ; c++){     // C: #input fmaps
  for(h=0; h<H ; h++){    // H: #rows in an fmap
   for(w=0; w<W ; w++){   // W: #cols in an fmap
    for(r=0; r<R ; r++){  // R: #rows in a filter
     for(s=0; s<S ; s++){ // S: #cols in a filter
        ofmap[k][h][w]+=ifmap[c][h+r][w+s] * weights[k][c][r][s];
}}}}}}
\end{lstlisting}

The loop order in Listing~\ref{lst:conv} can be written as $k\T c \T h \T w \T r \T s$, where, the operator $\T$ shows
the nesting order.  Due to the restricted memory capacity of on-chip buffers,
we often group pixels into {\em tiles}. In this case, an example
notation for a tiled execution where the feature maps ({\em channels}) are tiled (rows and columns grouped) will be of the form, $k \T c
\T h_T \T w_T  \T r \T s \T h \T w$;   the subscript $_T$ indicates the tile number, and the iterator
without a subscript retains its previous meaning (pointing to a single element). 
Figure~\ref{fig:convolution} shows a generic example, where $k$, $c$, $h$, and $w$ are tiled. We will follow a
consistent
terminology for all iterators, e.g., $C$ is the number of \ifmaps (input channels), $C_T$ is the size of a channel tile,
$c_T$ is the iterator for a channel tile, and $c$ is the iterator of a channel (see Table~\ref{tab:confi})
. Note that unless specified otherwise, the term
{\em tile} will refer to a group of pixels.

\begin{figure}[!ht]
\begin{center}
\footnotesize

\includegraphics[width=0.97\columnwidth]{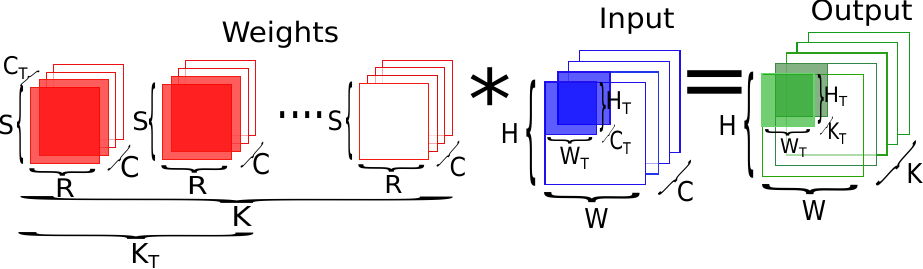}
\caption{Graphical representation of a tiled convolution operation. The \ifmaps, \ofmaps, 
input and output channels are tiled. \label{fig:convolution}}
\end{center}
\end{figure}

\subsection{TNPU and GuardNN}
We shall compare \fname with two state-of-the-art proposals: GuardNN~\cite{guardnn} and TNPU~\cite{tnpu}. A brief
description follows (see Section~\ref{sec:RelatedWork} for more details).

In GuardNN, every tile is associated with a version number (VN), which is incremented on every memory
write. The VNs are managed by the scheduler that runs on the host CPU, which has access to a TEE. MACs are generated
and maintained per memory block. A variant of GuardNN advocates for a larger block size (512 B); however, that
unnecessarily constrains the subsequent layer to read data in that order, which we found to be impractical for modern
CNNs where dataflow patterns are different for each layer.
TNPU on the other hand, maintains VNs in a Tensor Table that is stored in the host CPU's secure memory. It is
protected by an integrity tree. MACs are maintained per block, and are stored in an on-chip MAC cache. Most
architectures in this space including these and \fname do not use the regular on-chip caches and instead read/write
directly to main memory.

\section{System Design and Threat Model}
\label{sec:ThreatModel}

We use the same high-level system design and threat model as previous works~\cite{tnpu,guardnn}.

\begin{figure}[!h]
    \centering
    \includegraphics[width=0.9\columnwidth]{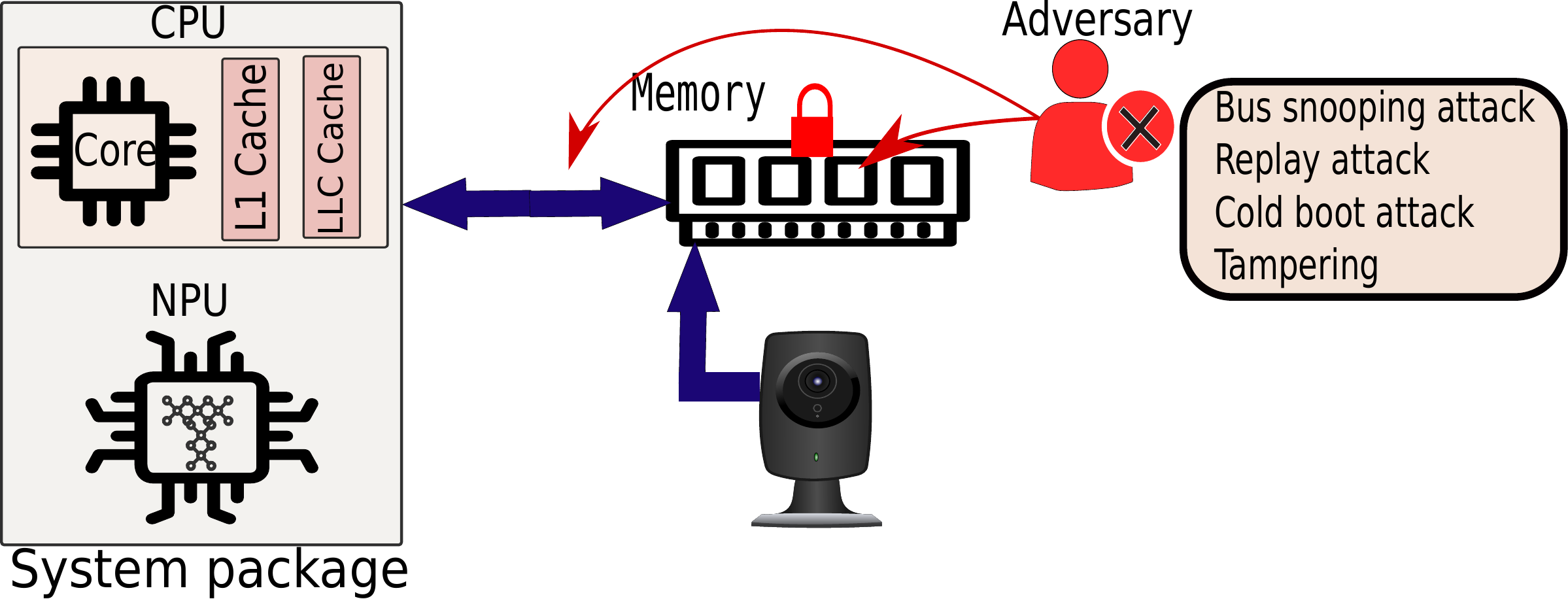}
    \caption{An overview of the system design and threat model}
    \label{fig:threat}
\end{figure}

A high-level overview of the system is shown in Figure~\ref{fig:threat} along with the possible attacks
that can be mounted. The CPU, NPU (\fname), and the caches
are in the same package (SoC or multi-chip module). The CPU can send instructions to the NPU directly via 
a secure channel; however,
the data transfer is through main memory. The CPU can optionally run a scheduler (possibly on a TEE) 
to coordinate the actions of
the NPU, or the NPU could completely take over after the CPU has sent it a few initialization instructions. 
Our design is unaffected by this choice. Insofar as the NPU is concerned, it needs to securely run an ML model,
layer by layer. We only focus on inferencing in this paper (like~\cite{tnpu}). 

The package comprising the CPU, NPU, and caches is within the Trusted Computing Base (TCB). A hacker can however
target the main memory and the CPU-memory bus (used by the NPU as well). A hacker in this case would include
the OS, hypervisor, malicious program, or even a person who has physical access to the main memory or memory bus.
She can try to read the data (eavesdropping), tamper with the data (integrity attack), replace the contents of
memory addresses with old data (replay attack), masquerade as a different entity (authentication attack), read
the memory address sequence from the memory bus and try to figure out the model parameters (model extraction attack 
(MEA)). Our base version of \fname protects against all types of aforementioned attacks other than MEA attacks. 
We shall show in Section~\ref{sec:scale} that it is possible to easily augment our architecture using ideas from Li et
al.~\cite{neurobfuscator} 
to also provide security against MEA attacks with a modest overhead. This version will also protect against
address-based side-channel attacks. However, we don't protect against power and EM side-channel based attacks.
Finally, like all prior work, we do not consider networks that have value-based pruning.

\section{Characterization of DNN Workloads in a Secure Environment}
\label{sec:characterization}

\begin{table}[!th]
\begin{center}
\footnotesize   
\begin{tabular}{c}

\begin{minipage}{0.99\columnwidth}
\begin{tabular}{|p{18mm}|p{15mm}||p{2cm}|c|}
\hline
\rowcolor{gray!20}
\multicolumn{4}{|c|}{Simulation configuration (similar to~\cite{tnpu})}\\
\rowcolor{white}
\hline
\textit{Parameter} & \textit{Value}  & \textit{Parameter} & \textit{Value}\\
\hline
PE array & $32\times32$ & Counter cache (secure NPU) & 4 KB \\ 
 Global buffer & 240 KB & MAC Cache (for secure NPU) & 8 KB\\  
  Frequency & 2.75 GHz  & Write mode & Write back \\
Dual-channel DRAM & DDR 4, 100 cyc (lat) &  Block Size & 64 B  \\
 \hline
 \hline
\end{tabular}
\vspace{3mm}
\end{minipage}

\\ 

\begin{minipage}{0.99\columnwidth}

\begin{tabular}{cc}

\begin{minipage}{0.64\columnwidth}
\begin{tabular}{|l|l|l|} 
\hline
\rowcolor{gray!20}
\multicolumn{3}{|c|}{ List of benchmarks (similar to~\cite{slalom,guardnn})} \\
\rowcolor{white}
\hline
\textit{Workload} & \textit{Layers} & \textit{Parameters} \\
\hline
MobileNet\cite{mobilenet} & 23 & 4.2 million \\
ResNet \cite{resnet} & 18 & 11 million  \\
AlexNet \cite{alexnet} & 13 & 62 million  \\
VGG16 \cite{vggnet} & 24 & 138 million \\
VGG19 \cite{vggnet} & 19 & 143 million\\
\hline
\end{tabular}
\end{minipage}
&
\begin{minipage}{0.33\columnwidth}
\begin{tabular}{|l|l|}
\hline
\rowcolor{gray!20}
\multicolumn{2}{|c|}{Terminology used} \\
\hline
Term & Meaning \\
\rowcolor{white}
\hline
$H$ & \# rows \\
$H_T$ & \# row tiles \\
$h_T$ & row tile \\
       & iterator \\
$h$ & row iterator \\
\hline
\end{tabular}
\end{minipage}
\end{tabular}
\end{minipage}

\end{tabular}

\caption{NPU configuration, list of benchmarks, and the
terminology used in the paper \label{tab:confi}}
\end{center}
\end{table}

\subsection{Setup and Benchmarks}
\label{sec:setup}

We characterize the behavior of popular benchmarks on an in-house cycle accurate CNN simulator that has been rigorously
validated with ARM SCALE-Sim~\cite{scalesim} and native hardware. It relies on a systolic array architecture to perform
convolutions. We relied on the Timeloop~\cite{timeloop} tool to provide the most optimal dataflow pattern. We show the
configuration of the simulated system in Table~\ref{tab:confi}. We
simulated a vanilla, \textit{unsecure} accelerator version as the {\em baseline}. The
workloads comprise popular neural
network benchmarks as shown in Table~\ref{tab:confi}. \textit{Layers} represents the total number of layers and
\textit{Parameters} represents the total number of tunable model parameters present in a specific benchmark. 
Additionally, we simulate a {\em secure} configuration that is quite similar to SGX-Client (see
Section~\ref{sec:sgxclient}).
The MACs and counter values are saved in the 8 KB MAC cache and 4 KB counter cache,
respectively (values taken from~\cite{tnpu}). All the models fit within the DRAM.
Finally, note that {\bf performance} is defined as the reciprocal 
of the simulated execution time (appropriately normalized).

\begin{figure}[!h]
   \centering
   \includegraphics[width=0.8\columnwidth]{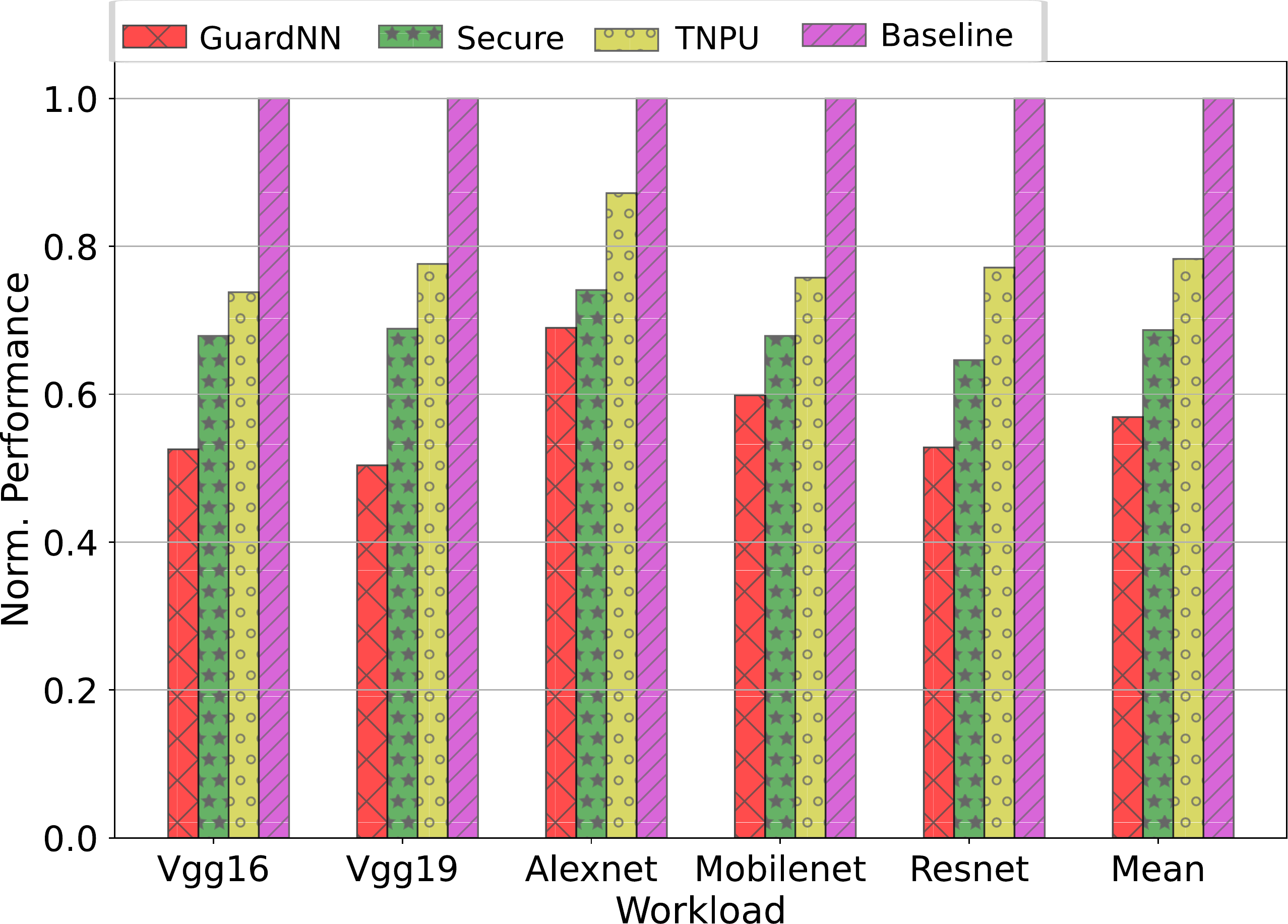}
   \caption{Performance comparison}
   \label{fig:latency}
\end{figure}  
\begin{figure}[!h]
    \centering
    \includegraphics[width=0.8\columnwidth]{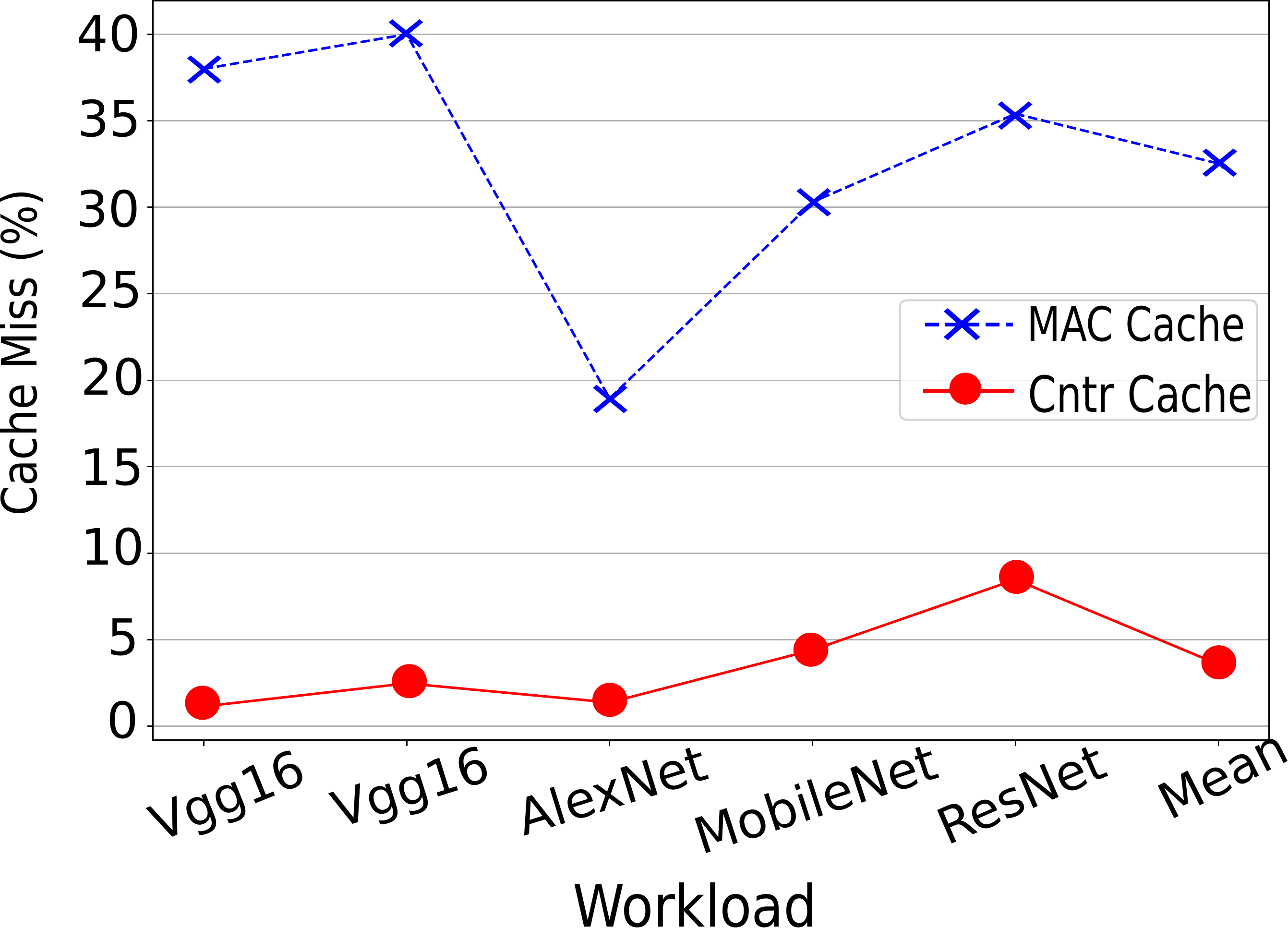}
    \caption{Cache miss rates for the MAC cache and counter cache, respectively.}
    \label{fig:mac_miss} 
\end{figure}  

\subsubsection{Performance Insights}   

Figure~\ref{fig:latency} shows the performance results. Secure and TNPU have an 8 KB MAC cache.
The secure configuration is 32\% slower than the baseline, TNPU is 22\% slower, and GuardNN is 44
\% slower.
Clearly, reducing the size of secure memory helps, and having a MAC cache also helps (as opposed to not having
one like in GuardNN). 

Figure~\ref{fig:mac_miss} shows the miss rates for the MAC cache and counter cache in the secure configuration.
We observe a high hit rate for the counter cache and a relatively lower hit rate for the MAC cache. 
Each 64-byte data block can store 16 four-byte pixels (element in a feature map). 
Each page of 64 blocks can store $64\times16$ pixels implying that a counter cache entry can keep
track of $64\times16=1024$ pixels. For uniform streaming data, after a cache miss, we may observe a cache hit for the
next 1024 consecutive pixels.
On the other hand, a 64-byte line can store only 8 MACs (each 8 bytes). 
Since each MAC protects a
block of 16 pixels (64 bytes), this means that a MAC cache block can 
track $16 \times 8=128$ pixels, which is $8 \times$ less
than a counter cache block. This means that the MAC cache miss rate will be much higher; however,
on the flip side its miss penalty will be much lower -- primarily a DRAM memory read as opposed to traversing a
Merkle tree for counters (DRAM+cache). In any case, streaming data such as pixels in DNNs have poor temporal locality
and the low absolute values of miss rates attest that. 

{\bf Conclusion: MAC caches have a very poor hit rate. Moreover, frequently
accessing secure memory to read VNs and MACs has a high overhead. Hence, we should try to avoid both.}

\begin{scriptsize}
\begin{table*}[!htb]
\footnotesize   
    \centering
    \begin{tabular}{||p{0.16\textwidth}|p{0.27\textwidth}||l|p{0.15\textwidth}|p{0.17\textwidth}||}
    \hline
    \rowcolor{gray!20}
    \textbf{Expression} & \textbf{Possible patterns} & \textbf{Expression} & \multicolumn{2}{|c}{\textbf{Possible
patterns}}\\
    \hline
   $[{1^{\alpha_K}, 2^{\alpha_K}...{\alpha_C}^{\alpha_K}}]^{\alpha_{HW}}$      &   \begin{minipage}{.19\textwidth}
  \includegraphics[scale=0.96]{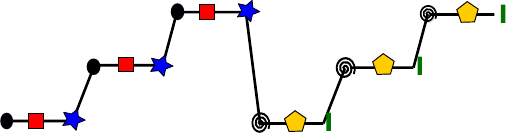}
    \end{minipage}
   & ${1^{\alpha_{K}\alpha_{HW}}, 2^{\alpha_K\alpha_{HW}}..\alpha_C^{\alpha_K\alpha_{HW}}}$     &  \begin{minipage}{.1\textwidth}
  \includegraphics[scale=1]{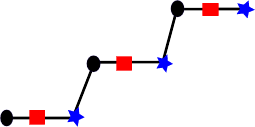}
    \end{minipage} &  \begin{minipage}{.1\textwidth}
  \includegraphics[scale=1]{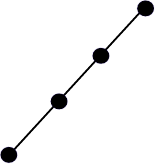}
    \end{minipage}  
    \\
    & \multicolumn{1}{l||}{P1:Multi-step} & &  P2:Step &  P3:Linear($\alpha_{K}\alpha_{HW}=1$)   \\
    
    \cline{2-2} \cline{5-5} 
    & \begin{minipage}{.05\textwidth}
  \includegraphics[scale=0.6]{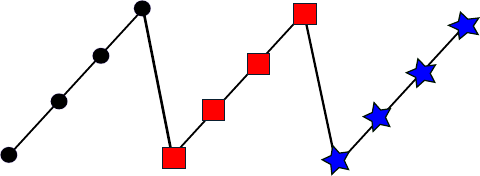} 
    \end{minipage} & $1^{\alpha_K},2^{\alpha_K}..{\alpha_C}^{\alpha_K}$ & 
 &
    \begin{minipage}{.1\textwidth}
  \includegraphics[scale=0.4]{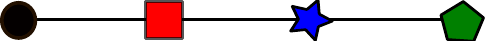}
    \end{minipage} \\
 & P4:Sawtooth ($\alpha_K=1$) & &  & P5:Line($\alpha_C=1$)   
    \\
     \hline
\end{tabular}

\begin{tabular}{ |p{0.23\linewidth}|p{0.13\linewidth}|p{0.27\linewidth}||p{0.13\linewidth}|p{0.15\linewidth}|}
\hline
\rowcolor{gray!20}
             & \multicolumn{2}{|c||}{\textbf{Input reuse}} & \multicolumn{2}{|c|}{\textbf{Output reuse}} \\
\cline{2-5}
\rowcolor{gray!20}
\textbf{Tiling style}  &  \textbf{Loop order}    & \textbf{Rd/Wr pattern} & \textbf{Loop order} &\textbf{ Rd/Wr pattern}\\
\hline
\rowcolor{blue!10}
\multicolumn{5}{|c|}{Tile movement along the channel}\\
\circled{1} Partial channel \cite{channelwise,multichannel}    & 
    $h_T\T w_T\T c \T k_T$
   & WP:   $[{1^{\alpha_K}, 2^{\alpha_K} \ldots {\alpha_C}^{\alpha_K}}]^{\alpha_{HW}}$
 & $h_T\T w_T \T k_T \T c$ & WP: $1^{\alpha_{K}\alpha_{HW}}$\\

\circled{2} Partial-multi-channel\cite{multichannel}  &  $h_T\T w_T\T c_T \T k_T$ & RP: $[{1^{\alpha_K}, 2^{\alpha_K}
\ldots {[\alpha_C-1]}^{\alpha_K}}]^{\alpha_{HW}}$
&  $h_T\T w_T \T k_T \T c_T$   &  RP:  --  \\
& & $Patterns$: P1,P2,P4 & & $Patterns$: P5\\
\hline

\rowcolor{blue!10}
\multicolumn{5}{|c|}{Tile movement along the width/height}\\
\circled{3} Partial channel\cite{romanet} &

$c\T h_T\T w_T \T k_T$
& WP:  
${1^{\alpha_{K}\alpha_{HW}}, 2^{\alpha_{K}\alpha_{HW}} \ldots \alpha_C^{\alpha_{K}\alpha_{HW}}}$ 
    &  -- & --  \\

\circled{4} Partial-multi-channel\cite{multichannel}   &   
 $c_T\T h_T\T w_T \T k_T$   &
RP: ${1^{\alpha_{K}\alpha_{HW}}, 2^{\alpha_{K}\alpha_{HW}} \ldots [\alpha_C-1]^{\alpha_{K}\alpha_{HW}}}$ 
    & -- & --\\
& & $Patterns$: P2,P3 & -- & -- \\
\hline
\hline

\circled{5} Channel-wise\cite{channelwise} & $ c \T k_T$;  
    & 
    WP: $1^{\alpha_K},2^{\alpha_K}..{\alpha_C}^{\alpha_K}$ & 
    $ k_T \T c$    
    & 
    WP: $1^{\alpha_K}$   \\

    & $ c_T \T k_T$; &
RP:  $1^{\alpha_K},2^{\alpha_K}..[{\alpha_C-1}]^{\alpha_K}$
        & $ k_T \T c_T$  &  RP: -- \\
        & & $Patterns$: P2,P3 & & $Patterns$: P5\\
\hline
\hline

\circled{6} Full-channel\cite{deep} &  $h_T \T w_T \T  k_T$ &  WP: $1^{\alpha_{K}\alpha_{HW}}$ & $h_T\T w_T \T k_T $
& WP: $1^{\alpha_K\alpha_{HW}}$  \\
&  & RP: -- & & RP: -- \\ 
 & & $Patterns$:P5 & & $Patterns$:P5 \\
\hline

\end{tabular}
\caption{\label{tab:dataflow_convolution} Pattern table for convolution: various possibilities for 
scheduling the input tiles for input reuse and output reuse. 
$\alpha_K = \frac{K}{K_T};\alpha_C = \frac{C}{C_T};\alpha_{HW} = \frac{H.W}{H_T.W_T};$ RP $\rightarrow$ Read pattern, WP
$\rightarrow$ Write pattern, $-$ refers to empty or not applicable}
\end{table*}
\end{scriptsize}

\section{Analytical Characterization of Patterns in ML Workloads}
\label{sec:PatternAnalysis}

Imagine a pair of observers sitting at the global buffer (GB) of the NPU and recording the version number of every tile
that is read or written. Let us refer to them as the read-observer and write-observer, respectively. For a layer,
they will see a sequence of VNs.  Prior works have used a tabular data structure to store the latest VN for each tile
(just before it is written to main memory). The same VN needs to be fetched from the table when the tile is read the
next time. We argue in this section that using a conventional table is an overkill. An ML application  has a very
well-defined behavior, and the sequence of VNs seen by the observers can be very nicely characterized (depending upon
the type of data reuse and feature map). We can thus supplant the multi-KB table stored in the host CPU's secure memory
with a simple mathematical formula processor that can generate VNs at runtime. This will be a part of the NPU, and the
storage requirement will be limited to a few registers.

\subsection{Convolution Layer}
In convolution, there are three possible types of data reuse: input reuse, weight reuse, and output reuse
\cite{romanet}. In all three cases, the output feature maps are updated, hence, let us focus on them.  

\subsubsection{Input Reuse (IR)}
The primary goal of this scheme is to minimize the number of times an \ifmap is accessed.  The stages are as follows:
\circlenew{1} The tiled \ifmaps are loaded into the GB. \circlenew{2} The \ifmaps are entirely reused in order to
compute the corresponding \ofmaps. \circlenew{3} The partially computed \ofmaps are stored back to memory.  They are
retrieved for the next \ifmap and updated. Let us look at various patterns generated by scheduling
\ifmap tiles in different ways~\cite{romanet}. The following text heavily refers to Table~\ref{tab:dataflow_convolution}
that compiles the most popular data flow patterns.
\vspace{2mm}

 \noindent \underline{Rows \circled{1} and \circled{2}  $\blacktriangleright$ \textit{Partial (multi) channel loading: }} 
Consider the first entry's loop order with $K_T=1$:  $h_T \T w_T \T c \T k$. We do not show the rest of the iterators because
it is assumed that once the data corresponding to a set of input/output tiles is fetched (as per the loop order), the
rest of the processing happens within the NPU. We are not concerned with internal aspects of the NPU, we only care
about the accesses to main memory. Now, as per this loop order, we consider an input tile, and then we compute the
results for a set of \ofmaps. We cycle through the \ofmaps, and then move to the next \ifmap.

Assume, $C=2$, $K=3$, and the GB can hold only one \ofmap tile at any point of time.
Each \ifmap tile is reused by three filters ($K=3$) to generate 3 distinct \ofmap tiles. 
Each \ofmap tile is associated
with a VN (incremented by 1 when the \ofmap is evicted from the GB). 
After the \ifmap tile has been fully utilized, all \ofmap tiles will have the same VN, as they are in the same
computational stage (only one \ifmap channel is processed). This leads to a VN write pattern $1^K$
as seen by the Write-observer. If we group a set of $K_T$ output tiles and consider them together, then 
this expression will change to $1^\alpha_K$, where $\alpha_K= K/K_T$. 

Thereafter, the next \ifmap tile from the next channel is read
from memory. These \ofmap tiles are again
sequentially updated -- we increment the VN of all the tiles from 1 to 2 (as they get evicted from the GB)
leading to the following pattern: $1^K 2^K$. The operation will repeat until all input channels are
processed.  After processing all the channels of an \ifmap tile, we schedule the input tile movement along the $w$-axis and
the aforementioned process will repeat
until all  $\alpha_{HW} = \frac{H\times W}{H_T \times W_T}$ \ifmap tiles are processed. The VN updates will
generate a simple deterministic pattern: Multi-step or Sawtooth . Additionally, the \ofmap
tile's \textit{read pattern} will be mostly identical to the write pattern.  The only difference is that we will not read the
final \ofmap. The final \ofmap will be read in the subsequent layer. 
In Row \circled{2}, we consider a group of $C_T$ channels together. The expressions are similar.

\vspace{2mm} 
\noindent \underline{Rows \circled{3} and \circled{4} $\blacktriangleright$ \textit{Movement along the Width/Height: }}
Consider the loop order, $c \T h_T \T w_T \T k_T$. Basically, we first cover an \ifmap and then move to the next. 
This means  
 $\alpha_{HW}=(H\times W)/(H_T \times W_T)$ \ifmap tiles will be accessed one after the other.
Each \ifmap tile will thus partially generate all the $K$
\ofmaps (in groups of $K_T$). All these
groups of \ofmap tiles will have the same VN.  Then we shall
move to the next \ifmap to update all the partially computed output tiles. We increment the VNs accordingly. The
process will repeat until all the \ofmap tiles are fully computed. 

\vspace{2mm} 
\noindent \underline{Rows \circled{5} $\blacktriangleright$ \textit{Channel-wise: }}

In this scheme, a single input channel
(channel-wise) or a group of input channels (multi-channel) of size $H \times W$ constitute an input tile. Due to the
stationary nature of the input tile, it will be reused $\alpha_k = K/K_T$ times to generate all of the $
\alpha_k $ \ofmap tiles of size $H \times W$ leading to the pattern ($1^{\alpha_{K}}$). 
These partially computed output tiles are updated when a new \ifmap tile is fetched from memory. 
The operation will be
repeated until all input channels are processed. 

\vspace{2mm} 
\noindent \underline{Rows \circled{6} $\blacktriangleright$ \textit{Full Channel: }}

This is a simple scheme in which all the channels required for the
generation of an output tile are available. The VN for a tile will be updated only once due to the availability of all
the
channels, leading to the pattern $1^{\alpha_{K}\alpha_{HW}}$.

\subsubsection{Output Reuse (OR)}
An output tile is completely computed in this scheme before being sent to memory.  Initially, \circlenew{1} successive
channels of the same \ifmap are loaded into the GB.  \circlenew{2} The partial \ofmap tile is reused until it is
fully computed.  The write patterns are very simple because there is no write-read-update cycle. They are of the form
$1^x$. There is no read pattern because partially computed \ofmap data is never read back.

\vspace{2mm}
\noindent \underline{Rows \circled{1} and \circled{2} $\blacktriangleright$ \textit{Partial(multi) channel loading}:
} Prior to storing an output tile in memory, it must be completely computed.  The VN of an output tile of size $H_T
\times W_T$ remains the same as it does not leave the GB. Then we proceed to the next output tile.  The
pattern generated above will repeat for $(H \times W)/(H_T\times W_T)$ times for each of the $(H \times
W)/(H_T\times W_T)$ tiles.

\begin{scriptsize}
\begin{table} [!h]
\footnotesize   
\begin{center}
\begin{tabular}{ |p{20mm}|p{18mm}|p{36mm}| } 
\hline
\rowcolor{gray!20}
\textbf{Tiling style} &  \textbf{Loop order}  & \textbf{Pattern}\\
 \hline
 \rowcolor{blue!20}
 \multicolumn{3}{|c|}{Filter-wise movement}\\
  \hline
\noindent \circled{1} Multi-channel wise \cite{romanet}  & $c_T \T k_T$ &
     
WP:${1^{\alpha_K},2^{\alpha_K} \ldots \alpha_C ^{\alpha_K}}$;
RP:${1^{\alpha_K},2^{\alpha_K} \ldots (\alpha_C -1)^{\alpha_K}}$
$Patterns$: P2,P3  \\
\hline

 \rowcolor{blue!20}
 \multicolumn{3}{|c|}{Channel-wise movement}\\
  \hline
\circled{2} Channel- & $k_T\T c$ &
WP: $1^{\alpha_K}$; RP: $-$ \\
 wise \cite{romanet}  &  & $Patterns$: P5 \\
\hline
\hline
\circled{3} Full  & $k_T$ &    
   WP: $1^{\alpha_K}$; RP: $-$ \\
 -filter\cite{deep} &   &    $Patterns$: P5 \\
\hline    

\end{tabular}
\caption{\label{tab:dataflow_weight} Pattern table for convolution: Different methods of
scheduling weight tiles for weight reuse. }
\end{center}
\end{table}
\end{scriptsize}

\subsubsection{Weight Reuse}
The steps involved in this process are as follows. Consider the first row.  \circlenew{1} The tiled weight matrix (4D tensor)
 of size $C_T \times K_T \times R
\times S$ is loaded in memory. \circlenew{2} $C_T$ \ifmaps of dimension $H \times W$ are loaded in the global buffer
\cite{deep}. \circlenew{3} The weights are reused to compute $K_T$ \ofmaps of dimension $H \times W$. The pattern
generation methodology is similar to the previous schemes. The generated patterns for all the rows are shown in Table
\ref{tab:dataflow_weight}.

\subsection{Other Kinds of Layers}
Let us now look at some other kinds of layers in DNNs. Most layers are extensions of simple convolution such as
the Fully Connected layer. 

\noindent \textit{Generative-Adversarial Networks (GANs)}: 
GANs \cite{gans} are composed of a discriminator and a generator network. To generate fake images, the generator uses
deconvolution (transposed/dilated), whereas the discriminator uses convolution to discern between fake and real images.
The pattern generation approaches for general convolution presented in Table \ref{tab:dataflow_convolution} will work
for any kind of convolution including deconvolution. To convert the input to the needed form, we may need to do some
pre-processing. 

\noindent \textit{Matrix Multiplication}:
We also analyze the data access pattern in the case of matrix multiplications as it is extensively used in several ML
workloads including transformers \cite{transformer}. We classify the different tiling scenarios and present our findings
in Table \ref{tab:dataflow_matrix}.

\begin{scriptsize}
\begin{table}[!h]
\footnotesize   
\begin{tabular}{ |p{0.17\linewidth}|p{0.18\linewidth}|p{0.5\linewidth}|} 
\hline
\rowcolor{gray!20}
\textbf{Tiling Style}   & \textbf{Loop order}  & \textbf{Pattern}\\
 \hline

\hline
\circled{1} Fix $P$  &
    $h_T \T c_T \T w_T$
  &  WP: $(1^{\alpha_{W}},2^{\alpha_{W}} \ldots \alpha_C ^{\alpha_{W}})^{\alpha_H}$  \\
\cite{matrix}  & & RP: $(1^{\alpha_{W}},2^{\alpha_{W}} \ldots (\alpha_C -1)^{\alpha_{W}})^{\alpha_H}$ \\

\hline
\hline

\circled{2} Fix $Q$ &
    $ c_T \T w_T \T h_T$
  &  WP: $(1^{\alpha_{H}},2^{\alpha_{H}}\ldots \alpha_C^{\alpha_{H}})^{\alpha_W}$ 
       \\
\cite{matrix}  & & RP: $(1^{\alpha_{H}},2^{\alpha_{H}} \ldots (\alpha_C-1)^{\alpha_{H}})^{\alpha_W}$\\
\hline

\hline
\hline
\circled{3} Fix $R$   &
   $w_T \T h_T \T c_T $
    &
    WP: $1^{\alpha_{HW}}$ \\
\cite{matrix}  & & RP: $-$ \\
\hline
\hline

\end{tabular}
\caption{\label{tab:dataflow_matrix} Pattern table for matrix multiplication ($R=P \times Q$): 
Various methods for tiling the input. Dimensions of $P$: $H \times C$, $Q$: $C \times W$ 
$\alpha_C = C/C_T; \alpha_H = H/H_T; \alpha_W  = W/W_T$
}

\end{table}
\end{scriptsize}

\subsubsection{Image Pre-processing/ Pooling}
Numerous machine learning applications require the input image to be in a specific format.
Additionally, it may be necessary to enhance an image's features prior to computation.
This necessitates an analysis of the data movement patterns generated by various image pre-processing methods.

We divide image pre-processing applications into three computation styles. The output channel is
sometimes completely dependent on a single input channel, the scenario is represented as \texttt{Style-1} $S_x=T_x(X)$, where $T_X$
represents the {\em transformation function}, and $X$ represents the input element.  Because there is no requirement to store
partially computed outputs in memory, the output access pattern will be linear.  However, as illustrated in Table
\ref{tab:dataflow_pool} (in the Appendix), the number of output tiles will vary.  We observe that pooling and \texttt{Style-1} computations follow
a similar pattern.  Typically, a window is positioned above the image to perform computations relative to the
surrounding pixels in order to generate an output pixel. \texttt{Style-2} depicts a scenario in which all input channels are
merged to form a single output channel $S=T(R,G,B)$, whereas \texttt{Style-3} depicts a scenario in 
which all input channels are merged
using various transformations to form multiple output channels. Please refer to 
Table~\ref{tab:style2} and Table~\ref{tab:style3} (in the Appendix). 

\begin{scriptsize}
\begin{table}[!ht]
    \centering
    \begin{tabular}{|p{0.99\columnwidth}|}
     \hlineB{2}
          \rowcolor{gray!10}
		  \textbf{Insight:} We note that the pattern of VN updates is highly deterministic and is quite similar across a range
of operations. All of the patterns can be expressed using a single master equation: 
$(1^\eta, 2^\eta \ldots \kappa ^\eta)^\rho$. The triplet $\langle \eta, \kappa, \rho \rangle$.  \\
          \hlineB{2}
    \end{tabular}
   
    \label{tab:my_label}
\end{table}
\end{scriptsize}

\section{Design of Seculator}
\label{sec:architecture}
 
\begin{figure}[!h]
    \centering
    \includegraphics[scale=0.34]{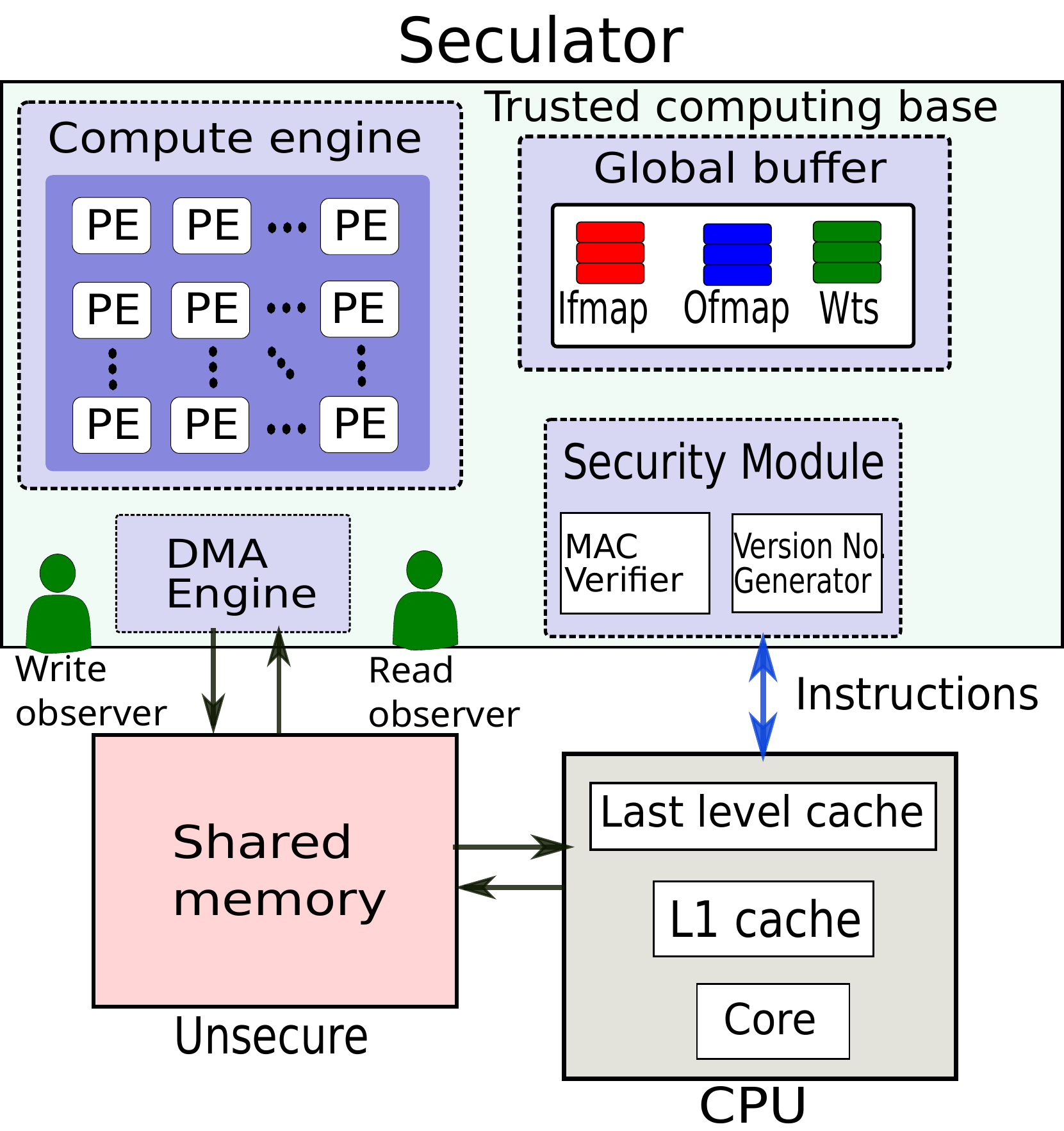}
    \caption{The high-level design of Seculator}
    \label{fig:acce} 
\end{figure}

\subsection{Overview}
The Seculator NPU architecture provides a hardware-assisted secure environment for the execution of neural networks.  A
high-level design of the framework is presented in Figure \ref{fig:acce}.  The host CPU securely delivers instructions
(using a shared key) to the
accelerator via a PCIe link to execute a layer of the CNN.  It also points to the location of the encrypted data
stored in memory.  Thereafter, the accelerator starts the execution.  Initially, all the model parameters and the
user input are
encrypted and stored in memory.  During the processing, the data is transferred to the NPU 
to perform the convolution operation. The compute engine is made up of a PE array and some local storage
(CE). When a layer is completed successfully, the accelerator notifies the host and writes the encrypted data to memory.
In the case of a security breach, a system reboot is performed. 

A security breach is detected by the security module, which will be in charge of securing the data and protecting it
from various threats. To reduce the overheads and memory traffic, we integrate the version number generator with the
security module such that the processed versions can be directly consumed. We explain the modules in the following
subsections.

\subsection{A Deterministic Version Generation Scheme}

\fname encrypts the output data when the data is evicted from the global buffer to memory. To automatically generate
the VNs for the output data, we thoroughly analyzed the movement of \textit{ofmap tile} data in Section
\ref{sec:PatternAnalysis}. An automatic VN generation scheme helps reduce the overheads associated with the storage and management of
VNs. Even though the storage requirements per se are not very high (max: 8 KB in prior work~\cite{tnpu}), the additional
complexity in the design and runtime for reading this table, which is stored in the host CPU's secure memory, is quite
onerous.
\fname overcomes this limitation by generating VNs automatically (once for every group of tiles)
 as per the master equation in Section~\ref{sec:PatternAnalysis}: 
$(1^\eta, 2^\eta \ldots \kappa ^\eta)^\rho$. 

The triplet $\langle \eta, \kappa, \rho \rangle$ is securely shared with the accelerator by the host CPU
along with a session-specific
encryption key (to decrypt data for the first time). Subsequently, we generate VNs based on memory accesses and the
value of the triplet. 

\subsection{Details of the Encryption Process}
We rely on AES counter-mode encryption (CTR) for encrypting each data  block. To encrypt a 64-byte data block, we employ
four parallel AES-128 engines. The 128-bit key is created by concatenating the accelerator's secret id (embedded within
it) with a random number
generated prior to execution. This technique ensures that the key is hardware-specific and changes with each execution. The
major counter value is created by concatenating the \fmap ID and layer ID, whereas the minor counter value is created by
concatenating the VN and index of the block within the \fmap.  This approach ensures that the counter value changes in
accordance with the index of the block in an \fmap (the same value is encrypted differently). 
The counters are encrypted using the AES-CTR mode to generate a
one-time pad (OTP). This OTP is XORed with the block data to create the ciphertext (standard algorithm).

\subsection{MAC Generation}
Unlike competing work, we do not maintain per-block MACs, we operate at the layer-level.
The reason that prior work  maintained per-block MACs is because they wanted to give the freedom
to subsequent layers to read data randomly (no pre-specified order). We also give the same freedom,
with one caveat, which is that all the \ofmaps that have been produced in Layer $i$ need to be
accessed at least once in Layer $i+1$. This is a very reasonable restriction. 

When we read/write a block, we compute its MAC. 
The 32-byte MAC is computed as $MAC=SHA_{256}(P||L||F||VN||I||B)$,
where, $P$ is the secret id of the accelerator, $L$ is the layer ID, $F$ is the id of the \fmap, 
$I$ is the block index within the
\fmap, and $B$ is the contents of the data block (64-bytes). $||$ is the concatenation operator. 

Let the MAC of a block that is being read/written be denoted by $MAC_B$. We maintain two 256-bit registers:
$MAC_R$ (for reads) and $MAC_W$ (for writes). For a block that is read we compute $MAC_R = MAC_R \oplus MAC_B$,
where $\oplus$ is the XOR operator. We do the same in the case of writes, albeit with the register $MAC_W$. 
As per Bellare et al.~\cite{xormac}, this scheme is quite secure (theoretically similar to chaining). 

We need to verify that whatever has been written is also read back without tampering. $MAC_W$ embodies everything that
 has been written.
If we analyze all the access patterns in Section~\ref{sec:PatternAnalysis}, we observe that in the same layer we
read everything back other than data written in the last iteration. This is read back in the next layer. 
In the next layer, we use one more register $MAC_{FR}$ (first read) that computes a MAC of all the \ifmap
data (\ofmaps of the previous layer) that is read for the first time. We use the layer id of the previous layer. 
Note that it is very easy to design a circuit using our master equation to figure out when an input tile
is read for the first time (not shown due to a lack of space).

The crucial condition that needs to hold is as follows. This is
provable from the equations shown in Section~\ref{sec:PatternAnalysis}. 

\begin{equation}
\label{eqn:firstread}
MAC_W = MAC_{FR} \oplus MAC_R 
\end{equation}

We can use two pairs of these registers (that alternate across layers) because of the overlaps in terms of usage
(need $MAC_R$ for the previous layer when the current layer is being processed). 

Let us now consider read-only data such as inputs and filter weights. Their VN remains the same (it is equal to the
last-generated VN in the previous layer for \ifmaps and 1 for filter weights). Without loss of generality, 
consider \ifmap data.
We maintain a separate $MAC_{IR}$ register for it.
If the same \ifmap tile is read an even number of times, the result of all the XOR operations to update $MAC_{IR}$
should be zero, otherwise it will be equal
to the XOR of the MACs -- same as the first-read data ($MAC_{FR})$. This is because the outputs of the previous layer should be the inputs of
the current layer. In either case, the inputs are verified because we are verifying that the ``first-read'' data for the
inputs is correct in Equation~\ref{eqn:firstread}.

\section{Evaluation}\label{sec:results}
\subsection{Setup}
We showed the detailed simulation setup and list of benchmarks in Section~\ref{sec:characterization}.
We evaluate 5 designs as shown in Table \ref{tab:configurations}.  They are the baseline design (no security), {\em
secure}
design ({\em ClientSGX}), TNPU, GuardNN, and \fname. 
We implemented the hardware of the pattern generator circuit, the SHA-256 and AES circuits in Verilog. 
We used the Cadence Genus Tool to synthesize, place, and route
the design in a 28 nm technology (scaled to 8 nm using the results in~\cite{scaling}).
TNPU and {\em secure} use an 8 KB MAC cache. We conducted simulators for an augmented design, {\em Seculator+}, 
that protects against MEA and bus snooping attacks (details in Section~\ref{sec:scale}).

\begin{scriptsize}
\begin{table}[h]
\footnotesize   
    \centering
    \begin{tabular}{|l|p{0.16\linewidth}|p{0.16\linewidth}|p{0.148\linewidth}|p{0.08\linewidth}|}
    \hline
    \rowcolor{gray!20}
    \textbf{Configuration}     &  \textbf{Integrity (MAC)} &\textbf{Encryption (AES)} & \textbf{Anti-Replay} & \textbf{MEA}\\
    \hline
    \textit{Baseline} & $\times$ & $\times$ & $\times$ & $\times$ \\
    \textit{Secure} & per-block & CTR & Counters & $\times$ \\
    \textit{TNPU} & per-block & XTS & VN & $\times$ \\
    \textit{GuardNN} & per-block & CTR & VN & $\times$\\
\hline
    \textit{Seculator} & per-layer & CTR & VN & $\times$\\
    \textit{Seculator+} & per-layer & CTR & VN & \checkmark\\
    \hline
    \end{tabular}
    \caption{Simulated designs}
    \label{tab:configurations}
\end{table}
\end{scriptsize}

\subsection{Verilog Synthesis Results}

The overhead associated with the hardware structures is shown in Table~\ref{tab:overhead}. 
We incur a marginal area overhead of $4210\ \mu m^2$ (area) and a sub-mW power overhead.

\begin{table}[h]
    \centering
    \begin{tabular}{|l|l|l|}
    \hline
    \rowcolor{gray!20}
    \textbf{Module} & \textbf{Area ($\mu m^2$)} & \textbf{Power$(\mu W)$}\\
    \hline
    AES-128 & 3900  & 640\\
    SHA-256 & 270 & 40 \\
    VN generator & 40 & 4.4\\
    \hline
    \hline
     \rowcolor{gray!10}
    \textbf{Tool} & \multicolumn{2}{|c|}{Cadence RTL Compiler, 8 nm \cite{scaling}}\\
    \hline
    \end{tabular}
    \caption{Overhead associated with the h/w structures}
    \label{tab:overhead}
\end{table}

\begin{figure}[!h]
    \centering
    \includegraphics[scale=0.27]{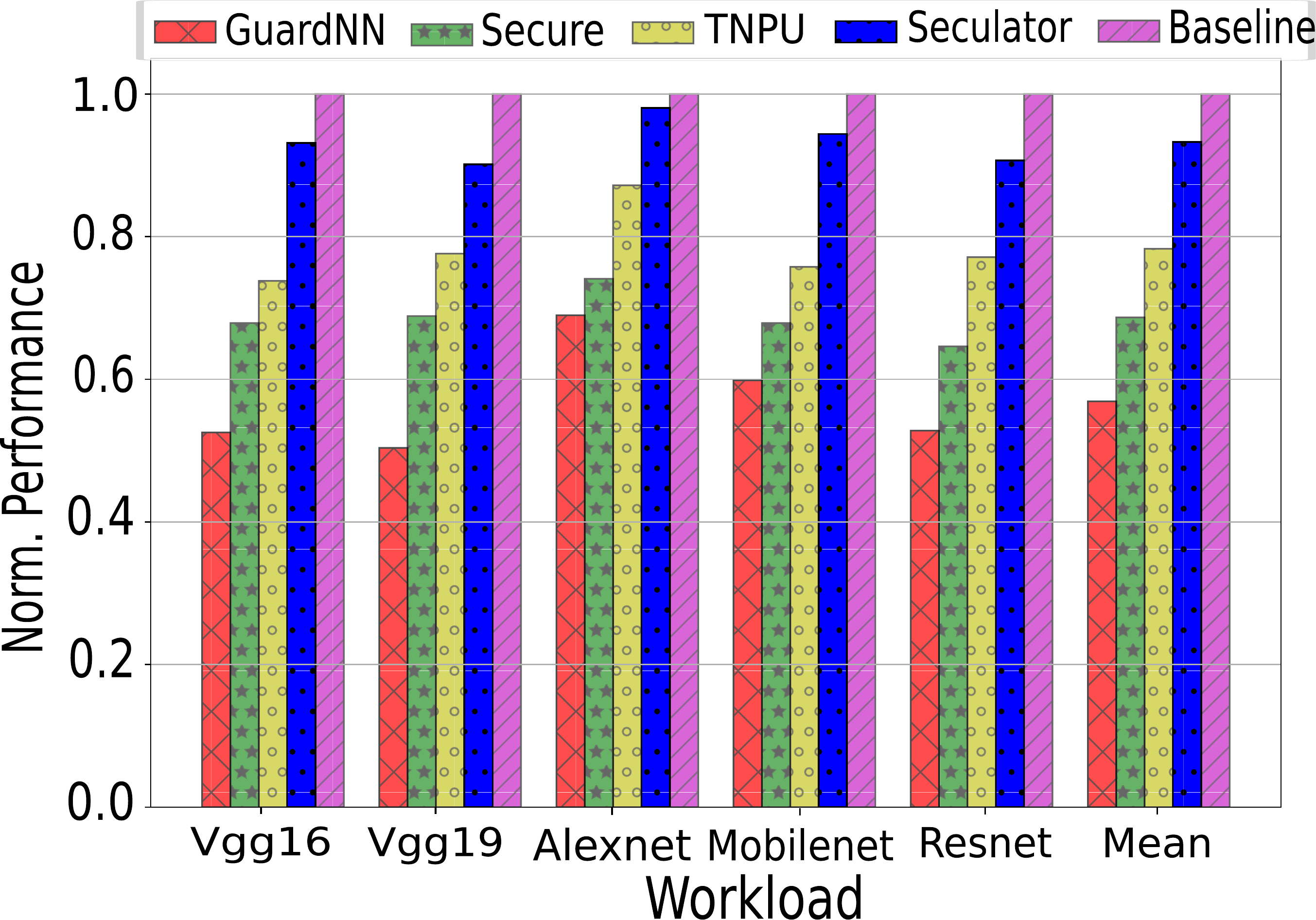}
    \caption{Normalized performance for different workloads}
    \label{fig:performance}
\end{figure}

\subsection{Performance Analysis}
Figure \ref{fig:performance} presents the performance results. The baseline configuration is used to normalize
the results.

We reduced the performance overhead by nearly $20\%$ compared to the state-of-the-art scheme TNPU. The fact that TNPU
verifies integrity block-by-block results in a large number of MAC accesses. With TNPU, we observed nearly $35\%$ misses
in the MAC cache since streaming data results in a high cache miss rate as described in
Section~\ref{sec:characterization}.  This
results in an increase in DRAM accesses as shown in Figure~\ref{fig:lat_access} (17\% more than \fname).  Additionally,
the access to the Tensor Table/tile stored in the secure memory location adds to the overhead. \fname simplifies the
design by computing on-chip versions, thus eliminating the requirement for additional DRAM accesses.
Because of direct DRAM accesses, there is a good correlation between the DRAM traffic and the performance.

In GuardNN, the MACs are generated according to the granularity of the accelerator data block (64 bytes). The scheme
does not use any MAC cache. All the MACs are stored in an off-chip memory. Each data read or write request is
accompanied by a MAC read/write request. This leads to a very high
memory traffic (40\% more than \fname) 
with a performance degradation of nearly $37\%$ as compared to \fname. 

\begin{figure}[!h]
   \centering
    \includegraphics[scale=0.3]{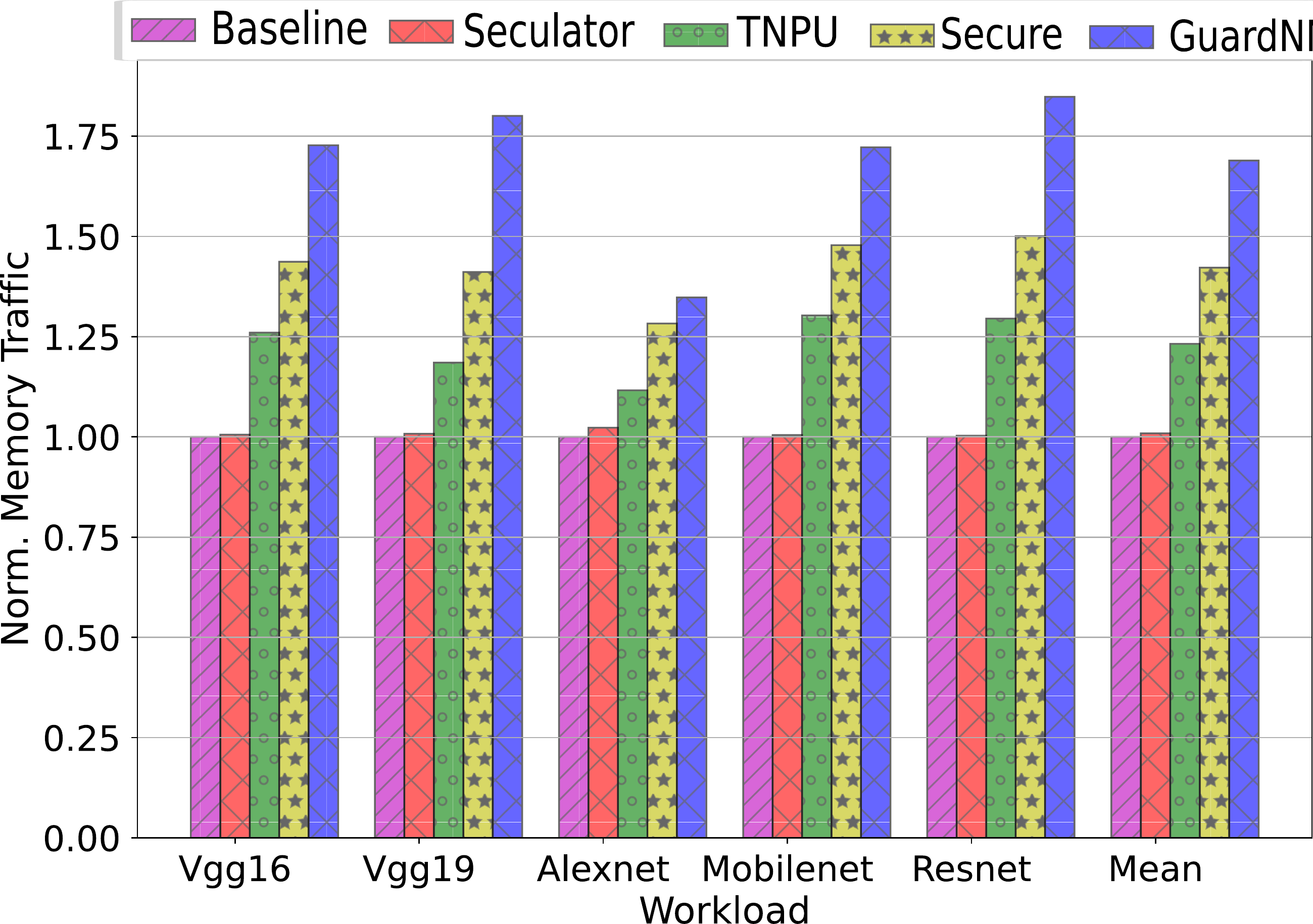}
    \caption{Normalized memory traffic for different workloads}
    \label{fig:lat_access}
\end{figure}

\subsection{Security Analysis}
\fname provides data confidentiality because all the data outside the TCB is encrypted. It provides security against
replay attacks because every 
memory write is assigned a different VN. It cannot be targeted by man-in-the-middle attacks because no other entity
has the secret id or the runtime key of the NPU. Finally, it provides integrity because we only read whatever is written. This holds
on a per-block basis and is not dependent on the {\em order of access} because we use the commutative XOR function and
include the block id (position of the block in the \fmap) in the MAC calculation. The cryptographic security guarantees
are quite similar to traditional chaining-based solutions~\cite{xormac}. It should also be noted that 
data blocks with the same content and address produce different ciphertexts over time (because of the VN and the
execute-time key). Finally, note that our order-independent
MACs can be visualized as hashing read and write sets to 32-byte values. 
From the master equation and the check in the subsequent layer, we can conclude that the sets are the same (a formal
proof not included for lack of space). Both the generated VNs and the MAC verification scheme were rigorously experimentally
validated. 

\begin{figure}[!h]
    \centering
    \includegraphics[scale=0.35]{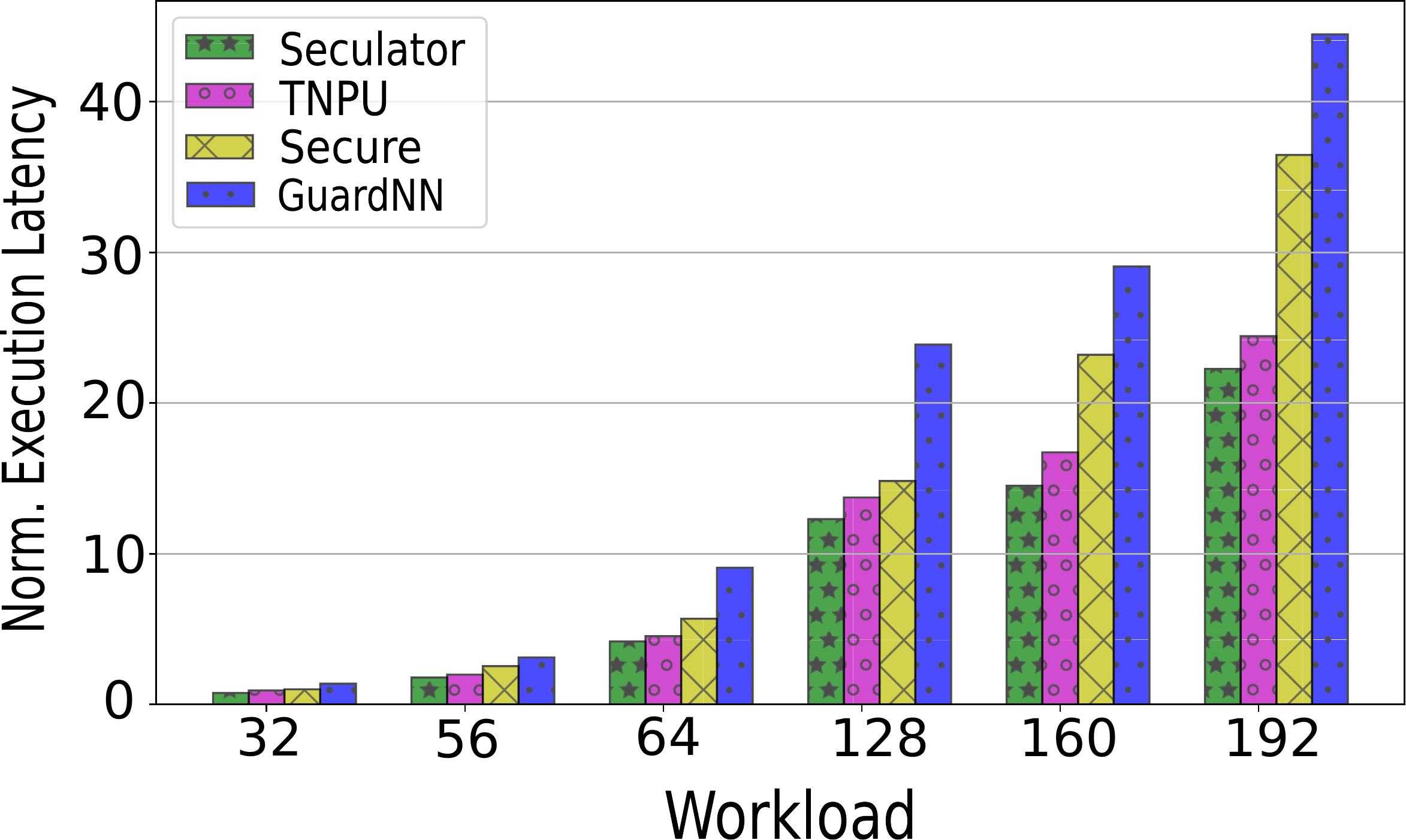}
    \caption{Normalized execution latencies for layer widening (normalized to $32\times32\times3$)}
    \label{fig:time_scale}
\end{figure}

\subsection{Scalability Analysis: Seculator+}
\label{sec:scale}
Li et al.~\cite{neurobfuscator} suggest many methods to prevent MEA and memory-based side-channel
attacks. A key technique is layer widening:
 increase the size of all layers by padding junk data such that it is not possible to find
their real size.  To investigate the
effect of layer widening, we increase the size of the base layer ($32\times32\times3$) to
($56\times56\times3),(64\times64\times3),(128\times128\times3),(160\times160\times3)$ and $(192\times192\times3)$ as
shown in Figure \ref{fig:time_scale}. We observe that \fname is the most scalable.

\begin{footnotesize}
\begin{table*}[!h]
    \begin{center}
    \footnotesize 
    \begin{tabular}{|p{0.16\textwidth}|p{0.1\textwidth}|p{0.11\textwidth}|p{0.1\textwidth}|p{0.07\textwidth}|p{0.15\textwidth}|p{0.06\textwidth}|p{0.1\textwidth}|}
    \hline
    \rowcolor{gray!20}
    \textbf{Work} & \textbf{Security} & \multicolumn{2}{|c|}{\textbf{Freshness(via VNs)}} &
\multicolumn{2}{|c|}{\textbf{Integrity(via MACs)}} & \textbf{MEA} & \textbf{Tile size} \\
    \cline{3-6}
    \rowcolor{gray!20}    
     & (via encryption) & Granularity & Space & Granularity & Space & \textbf{prot.} &\\ 
    \hline
    Outsourcing computations~\cite{slalom,darknight,hybridtee} & Partial & Block & $BTm + TM$ & Block & $BTH$ & $\times$ & \multirow{2}{*}{64 blocks}  \\
    \cline{1-7}
    SGX$+$ Optimizations~\cite{vessels, memory,occlumency}  & Full & Block & $BTm + TM$ & Block &  $BTH$ & $\times$ & \\
    \hline
    \rowcolor{gray!20}
    \multicolumn{8}{|c|}{Custom accelerators}\\
    \hline
    NPUFort~\cite{npufort} & Partial & Block  & $BTV$  & -- & -- & Obscure time* & \multirow{5}{1.2cm}{Depends on resources, scheduling scheme and workload} \\
     \cline{1-7}
    Seal~\cite{seal}  & Partial & Block &  $BTV$ & -- & -- & $\times$ &\\
    \cline{1-7}
    TNPU~\cite{tnpu} & Full & Tile  & $TV$  & Block  & $BTH$ (on-chip) & $\times$ &\\ 
    \cline{2-7}
    
    & \multicolumn{6}{c|}{VNs for each tile are stored in a secure memory (similar to SGX) \cellcolor{blue!10}}  &\\
    \cline{1-7}
    GuardNN~\cite{guardnn} & Full & Tile  & $TV$ &  Block  & $BTH$ (off-chip) & $\times$ &\\ 
    \cline{2-7}
    & \multicolumn{6}{c|}{VNs are stored and managed by a scheduler running on the host CPU (uses a TEE) \cellcolor{blue!10}} &\\
    \cline{1-7}
    \cline{1-7}
    \cline{1-7}
    \textbf{Seculator} &  \textbf{Full} & \textbf{Layer}  & \textit{{V}} & \textbf{Layer}  &  \textit{\textbf{O(H)
(on-chip)}} & \textbf{Widen layers} &\\
    \cline{2-7}
    & \multicolumn{6}{c|}{VNs are not stored, but generated using on-chip FSM \cellcolor{blue!10} } &\\

    \hline
    \multicolumn{8}{|c|}{$H$: MAC size, $V$: VN size, $M$: Major counter size, $m$: Minor counter size, $T$: Total \# of
tiles, $B$: \# of blocks per tile }\\

    \multicolumn{8}{|c|}{*Secure and unsecure fmaps have different latencies resulting in a fuzzy separation (in terms of temporal behavior)}\\
    \hline
     
    \end{tabular}
    \caption{\label{tab:relwork}A comparison of related work  }
    \end{center}
\end{table*} 
\end{footnotesize}

\section{Related Work}
\label{sec:RelatedWork}

TEEs provide a secure platform for the execution of CNN workloads even in an untrusted environment; additionally, they
outperform computationally heavy cryptographic methods such as fully homomorphic encryption~\cite{chet}. Recent
works focus on \circled{1} outsourcing the computation to an untrusted GPU \cite{slalom,darknight}; \circled{2}
optimizing a pre-existing TEE such as SGX or Tensorcone \cite{hybridtee};  and \circled{3} designing a secure, custom
accelerator~\cite{guardnn}. We present a brief comparison of related works in Table~\ref{tab:relwork}.

\subsection{Outsourcing Computations}
Slalom\cite{slalom} and DarkNight\cite{darknight} focus on providing a hybrid and secure execution platform to users.
They split the computations between a TEE and an untrusted GPU. 
Layers are obfuscated and delegated from the TEE to the GPU. The output from the unsecure GPU is verified using the
Freivald's algorithm \cite{slalom}.  The central insight behind the data obfuscation step is that convolution is a bilinear
operation. Instead of directly exposing the inputs to an untrusted third party, the CPU adds controlled noise. 
The bilinear property will ensure that the noise can later on be removed from the computed result. The overheads associated with 
input blinding, output verification along with the additional communication overhead due to data transfer from the TEE to
the GPU cannot be avoided. \fname relies on a TEE alone, hence, these overheads are not involved.  

HybridTEE\cite{hybridtee} divides the computation between a constrained-resource local TEE (ARM Trustzone) and a
resource-rich remote TEE (Intel SGX) based on the presence of security-critical features. A separate algorithm is used
to detect which parts of the image are security-critical (the SIFT and YOLO algorithms are used). However, executing an
auxiliary DNN to find security-critical features might lead to high overheads and may compromise system security.

\subsection{Optimizing Intel SGX}
Several works such as \cite{vessels, memory,occlumency} focus on optimizing the execution of the whole DNN within a TEE.
Vessels \cite{vessels} solves the problem of low memory re-usability and a high memory overhead by creating an optimized
\textit{memory pool} for the TEE. This is done by characterizing the memory usage patterns in a CNN layer. Similarly,
Truong et al. \cite{memory} propose to partition the layers to minimize the memory usage. In Occlumency
\cite{occlumency}, the authors proposed a memory-efficient feature map allocation technique and partitioning convolution
operation to optimize the CNN execution in SGX. 

On the contrary, DeepEnclave\cite{deepenclave} aims to secure only the user data. It optimizes the memory utilization by
executing the initial few layers inside the secure enclave and the latter outside. It is important to note that these
works use the vanilla version of client-side SGX where Merkle trees and eviction trees are maintained for the entire
data. However, we should note that such memory optimizations lose their utility in modern server-side SGX-based designs
where the enclave size can be as large as 512 GB \cite{xeon}. The key advantage of our scheme is that we provide 
all security guarantees without the additional overheads of counters and Merkle trees.

\subsection{Designing a Custom TEE}
GuardNN~\cite{guardnn} and TNPU~\cite{tnpu} 
focus on bringing security guarantees provided by traditional secure processors such as Intel SGX to
custom accelerators. The majority of the overheads in a conventional secure processor comes from cache misses and Merkle
tree traversals. GuardNN and TNPU both aim to eliminate the Merkle tree and counters with a more sophisticated and
efficient VN management mechanism that makes use of a DNN's extremely deterministic and statically defined memory access
patterns. 

\begin{table*}[!t]
\begin{center}
\footnotesize   
\begin{tabular}{c}

\begin{minipage}{\textwidth}
\begin{tabular}{cc}
\begin{minipage}{0.5\textwidth}
\begin{tabular}{ |p{0.29\linewidth}|p{0.18\linewidth}|p{0.17\linewidth}|} 
\hline
\rowcolor{gray!20}
\textbf{Tiling Style}  &  \textbf{Loop} & \textbf{Pattern} \\
\rowcolor{gray!20}
 & \textbf{order} & \textbf{\texttt{Style-1}} \\
\rowcolor{white}
 \hline
\circled{1} Channel-wise &
     $ k $ & WP:$1^{\alpha_K}$ \\

\circled{2} Multi-channel &  
    $k_T$ & RP:$-$\\
 
\hline
\hline

\circled{3} Partial channel &
   $h \T w \T k_T$   & WP:${1^{\alpha_K}}^{\alpha_{HW}}$\\

\circled{4} Partial-multi-channel  
     & $h_T \T w_T \T k_T$  & RP:$-$
      \\
\hline
\hline

\circled{5} Full-channel  & $h_T \T w_T$ & WP: $1^{\alpha_{HW}}$, RP: $-$   \\

\hline
\end{tabular}
\caption{\label{tab:dataflow_pool} Pattern table for image pre-processing (\texttt{Style-1})/Pooling $(S_x=T_x{X})$: Various
possibilities for scheduling the output tiles for performing pooling or image pre-processing $(H_T, W_T, C_T)$
represents the tiling factor. $(C=K)$ }
\end{minipage}

&

\begin{minipage}{0.5\textwidth}

\begin{tabular}{ |p{0.26\linewidth}|p{0.19\linewidth}|p{0.39\linewidth}||} 
\hline
\rowcolor{gray!20}
\textbf{Tiling Style}   & \textbf{Loop order} & \textbf{Pattern (\texttt{Style-2})}   \\
\rowcolor{white}
\hline
\circled{1} Channel-wise   &  $c$ & 
    WP: $1$, RP: 1
    \\
 \circled{2} Multi-channel   &  $c_T$ & 
    WP: $1$, RP: 1
    \\

\hline
\hline

\rowcolor{blue!10}
\multicolumn{3}{|c|}{Tile movement along the channel}\\
\rowcolor{white}
\circled{3} Partial channel  & $h_T \T w_T \T c$ 
   & 
    WP: $1^{\alpha_{HW}}$
    \\
\circled{4} Multi channel    & $h_T \T w_T \T c_T$ 
   & 
     RP:$-$
    \\

\rowcolor{blue!10}
\multicolumn{3}{|c|}{Tile movement along the width/height}\\
\rowcolor{white}
\circled{5} Partial channel & $c \T h_T \T w_T$ 
 &    WP: $1^{\alpha_{HW}},2^{\alpha_{HW}}\ldots \alpha_C ^{\alpha_{HW}}$
    \\
    
\circled{6}    Multi channel & $c_T \T h_T \T w_T$ 
 &     RP:$1^{\alpha_{HW}},2^{\alpha_{HW}}\ldots [\alpha_C-1]^{\alpha_{HW}}$\\

\hline
\hline

\circled{7} Full-channel &  
    $h_T\T w_T$ & WP: $1^{\alpha_{HW}}$, RP: $-$  \\

\hline
\end{tabular}
\caption{\label{tab:style2} Pattern table for image pre-processing (\texttt{Style-2}: $S = T(R,G,B)$): Different methods
of
scheduling input tiles (K=1)}

\end{minipage}
\end{tabular}
\end{minipage}

\\

\begin{minipage}{\textwidth}
\begin{tabular}{ |p{0.19\linewidth}|p{0.13\linewidth}|p{0.14\linewidth}||p{0.12\linewidth}|p{0.3\linewidth}|} 
\hline 
\rowcolor{gray!20}
\multicolumn{3}{|c|}{Output Reuse} & \multicolumn{2}{|c|}{Input Reuse}\\
\hline
\rowcolor{gray!20}
\textbf{Tiling Style} & \textbf{Loop order (OR)} & \textbf{Rd/Wr Pattern}  & \textbf{Loop order (IR)} & \textbf{Rd/Wr Pattern}  \\
\rowcolor{white}
 \hline
\circled{1} Channel-wise  & $c \T k_T$  &  WP:$1^{\alpha_K}$  & $k_T \T c$ & WP:$1^{\alpha_K},2^{\alpha_K}\ldots
\alpha_C^{\alpha_K}$   \\

\circled{2} Multi-channel & $c_T \T k_T$  & RP:$-$ & $k_T \T c_T$ & RP: $1^{\alpha_K},2^{\alpha_K}\ldots (\alpha_C
-1)^{\alpha_K}$\\

\hline
\hline

\rowcolor{blue!10}
\multicolumn{5}{|c|}{Tile movement along the channel}\\
\rowcolor{white}
\circled{3} Partial channel  &
     $h_T \T w_T \T k_T \T c$   & 
     WP:${1^{\alpha_{K}}}^{\alpha_{HW}}$  & 
    $k_T\T h_T \T w_T \T c$ & 
    WP:${(1^{\alpha_K},2^{\alpha_K}..\alpha_C ^{\alpha_K}})^{\alpha_{HW}}$ \\
\circled{4}  Multi channel   &
     $h_T \T w_T \T k_T \T c_T$ & 
     RP: $-$ &
     $k_T\T h_T \T w_T \T c_T$ & 
     RP:${(1^{\alpha_K},2^{\alpha_K} \ldots (\alpha_C -1)^{\alpha_K}})^{\alpha_{HW}}$
   
   \\

\rowcolor{blue!10}
\multicolumn{5}{|c|}{Tile movement along the  width/height}\\
\rowcolor{white}
\circled{5} Partial channel/Multi channel &  
    
  -- 
& -- & $k_T \T h_T \T w_T \T c$ & 
 WP: $1^{\alpha_{K}\alpha_{HW}},2^{\alpha_{K}\alpha_{HW}} \ldots \alpha_C ^{\alpha_{K}\alpha_{HW}}$

\\

\circled{6} Multi channel &  
    
  -- 
& -- & $k_T \T h_T \T w_T \T c_T$ & 
 
 RP:$1^{\alpha_{K}\alpha_{HW}},2^{\alpha_{K}\alpha_{HW}} \ldots (\alpha_C -1)^{\alpha_{K}\alpha_{HW}}$
 
\\

\hline
\hline

\circled{7} Full-channel   &  
    $h_T\T w_T\T k_T$
   & 
    WP: ${1^{\alpha_{k}}}^{\alpha_{HW}}$; RP: $-$
    & 
    $k_T \T h_T \T w_T$
    & WP: ${1^{\alpha_{HW}}}^{\alpha_K}$; RP: $-$\\

\hline
\end{tabular}
\caption{\label{tab:style3} Pattern table for image pre-processing (\texttt{Style-3}: $S_i = T_i(R,G,B))$
}
\end{minipage}
\end{tabular}

\end{center}
\end{table*}

TNPU needs a ``Tensor Table'' to keep track of the output tile updates (every update is associated with a new version
number). Since input tiles are never updated, the VNs linked with them are never updated. TNPU divides the secure memory
into two regions: the first is protected by the information stored in the Tensor Table (Region 1), while the second 
128 KB secure memory region is protected by a system similar to {\em SGX-Client} (Region 2).  The Tensor Table protects
Region 1, and the table itself is stored in Region 2.
MACs are generated at the granularity of individual blocks and they are stored in an
8 KB on-chip MAC cache (overflows go to main memory).

GuardNN relies on counter-mode encryption, which requires VNs.  For memory writes, the accelerator generates VNs using
on-chip counters, which are managed by the scheduler (running on the host CPU with a  secure TEE). 
For memory reads, VNs are received from the scheduler on the host CPU.
GuardNN also generates block-specific MACs which are not stored in an on-chip cache (read directly from main memory).

Due to the high memory requirements of DNNs, generating per-block MACs like GuardNN and TNPU results in high memory
overheads. Even if an on-chip cache is used, the situation does not improve significantly because caches are not
optimized for streaming data. \fname addresses these two concerns by significantly reducing the number of
MACs required to verify the integrity of a DNN. 
Additionally, because both GuardNN and TNPU are incapable of providing any additional security guarantees, both are
susceptible to \textit{model extraction attacks (MEAs)}, which rely on side-channel leaks such as addresses or memory
bus snooping. The {\em Seculator+} architecture is scalable, and can thus be used to implement layer widening
very efficiently. This will help protect against the aforementioned class of attacks. 

For GPUs, \cite{gpu2} and \cite{gpu1} provide a secure execution environment that is quite similar to GuardNN. 
They are not specific to neural networks and thus don't leverage their characteristics.
NPUFort~\cite{npufort} is a custom accelerator that encrypts security-critical features similar to HybridTEE.
We, on the other hand, focus on full encryption and our system is tailored for neural networks. 

\section{Conclusion}\label{sec:conclusion}
We showed that competing works such as GuardNN and TNPU naturally segue into the \fname proposal. It is a
logical successor in this line of work, where the move to layer-level security checks is complete. For
realizing this, it was necessary to mathematically characterize a large number of memory access patterns
(with different levels of stationarity) and encode them efficiently. The VN generator coupled with
the MAC verifier helped realize layer-level operations. Because of the reduced
need for data storage and consequently reduced DRAM traffic, it was possible to reduce wasted cycles, and
achieve an additional 16\% throughput as compared to TNPU, which is the nearest competing work. We could further
show that our mechanism could be tweaked to introduce controlled noise into the execution such that it
becomes harder to mount address snooping or side-channel attacks.

\begin{appendix}
We show the results for image pre-processing (\texttt{Style-1}, \texttt{Style-2}, and \texttt{Style-3})
 in Tables~\ref{tab:dataflow_pool},
\ref{tab:style2}, and \ref{tab:style3}.
\end{appendix}

\newpage


\bibliographystyle{IEEEtranS}
\bibliography{Ref}

\end{document}